# Measurement of Leakage Neutron Spectra for Tungsten with D-T Neutrons and Validation of Evaluated Nuclear Data

S. Zhang[a][1,2], Z. Chen[1,*], Y. Nie[3], R. Wada[1], X. Ruan[3], R. Han[1], X. Liu[1,2], W. Lin[1,2], J. Liu[1], F. Shi[1], P. Ren[1,2], G. Tian[1,2], F. Luo[1], J. Ren[3], J. Bao[3]

[1]Institute of Modern Physics, Chinese Academy of Sciences, Gansu, Lanzhou 730000, China

[2]University of Chinese Academy of Sciences, Beijing 100049, China

[3]Science and Technology on Nuclear Data Laboratory, China Institute of Atomic Energy, Beijing 102413, China

[*]Corresponding author. *E-mail address:* zqchen@impcas.ac.cn

**Abstract:** Integral neutronics experiments have been investigated at Institute of Modern Physics, Chinese Academy of Sciences (IMP, CAS) in order to validate evaluated nuclear data related to the design of Chinese Initiative Accelerator Driven Systems (CIADS). In present paper, the accuracy of evaluated nuclear data for Tungsten has been examined by comparing measured leakage neutron spectra with calculated ones. Leakage neutron spectra from the irradiation of D-T neutrons on Tungsten slab sample were experimentally measured at 60˚ and 120˚ by using a time-of-flight method. Theoretical calculations are carried out by Monte Carlo neutron transport code MCNP-4C with evaluated nuclear data of the ADS-2.0, ENDF/B-VII.0, ENDF/B-VII.1, JENDL-4.0 and CENDL-3.1 libraries. From the comparisons, it is found that the calculations with ADS-2.0 and ENDF/B-VII.1 give good agreements with the experiments in the whole energy regions at 60˚, while a large discrepancy is observed at 120˚ in the elastic scattering peak, caused by a slight difference in the oscillation pattern of the elastic angular distribution at angles larger than 20˚. However, the calculated spectra using data from ENDF/B-VII.0, JENDL-4.0 and CENDL-3.1 libraries showed larger discrepancies with the measured ones, especially around 8.5-13.5 MeV. Further studies are presented for these disagreements.

## 1. Introduction

China has started to develop the Chinese Initiative Accelerator Driven Systems (CIADS) project and is now underway vigorously. This project mainly aims for high radioactive nuclear waste transmutation, fuel breeding and clean energy production. In the design of CIADS, high intensity of ~1GeV proton beam bombards on heavy metal spallation target inside the subcritical reactor, and provides external neutron source for the subcritical reactor. The combination of evaluated nuclear data with a Monte Carlo transportation code like MCNP [1] is widely utilized for designing such kind of nuclear engineering facilities. However, the transportation codes and evaluated nuclear data used need to be validated through integral experiments [2-9].

Tungsten is proposed to be one of the most promising candidate spallation targets and other structural materials of the CIADS project, as well as an important material in fusion devices. There some integral experiments and evaluations for Tungsten [10-15] have been reported for benchmarking evaluated nuclear data related to the design of fusion devices. In this paper, the integral benchmark experiments for Tungsten at Institute of Modern Physics, Chinese Academy of Sciences (IMP, CAS) are presented. The experimental setup and data analyzing procedure are described in detail below. The measured results are compared with MCNP-4C [1] calculations using the evaluated nuclear data derived from the libraries of ADS-2.0 [16], which is a test library for Accelerator Driven Systems and new reactor design, ENDF/B-VII.0 [17], ENDF/B-VII.1 [18], JENDL-4.0 [19] and CENDL-3.1 [20]. The comparisons are made in both the spectrum shape and the calculation-to-experiment (C/E) ratio of the spectrum integrated over five energy regions.

## 2. Experiment
## 2.1 Experimental setup

The measurement was performed by using the integral experiment facility of China Institute of Atomic Energy (CIAE). A schematic view of the experimental arrangement is shown in Fig. 1. Deuteron ($D^+$) beam was accelerated by the 400 kV Cockcroft-Walton accelerator and bombarded on the Tritium-Titanium (T-Ti) target to produce neutrons by the T (d, n) $^4$He fusion reactions. The $D^+$ beam was bunched about 2.5 ns width in FWHM (Full Width at Half Maximum) and the repetition rate was 1.5 MHz. The average beam current and beam energy were about 20 uA and 300 keV, respectively. The neutron energy generated in the experiment was about 14.8 MeV in the forward direction. For monitoring the intensity of source neutrons, an Au-Si surface barrier semiconductor detector (Silicon counter) was positioned at 135° with respect to the deuteron beam by counting the associated $^4$He particles. A stilbene scintillation crystal (Monitor) of 5.08 cm in diameter and 5.08 cm in length was also placed at about 8 m from the target at incident beam direction for monitoring the source neutrons. Another BC501A liquid scintillator counter (Neutron detector), which was located behind the concrete wall with a collimated hole, had a size of 5.08 cm in diameter and 2.54 cm in length for detecting the leakage neutrons from sample. The detection angle can be changed by moving the sample along the dashed line in Fig. 1. A Tungsten slab sample in size of 10 ×10 ×7 cm$^3$ was used in the present experiment. A collimator system, which was made up of iron, polyethylene, and lead, was placed between the sample and the neutron detector to reduce background neutron. The aperture size of this collimator was determined so that the whole surface of the target sample can be viewed by the neutron detector. The 90 cm long shadow bar of Cu was used to block the direct injection from the source neutrons. Using such a heavy shielding and collimating system, high foreground to background ratio was achieved.

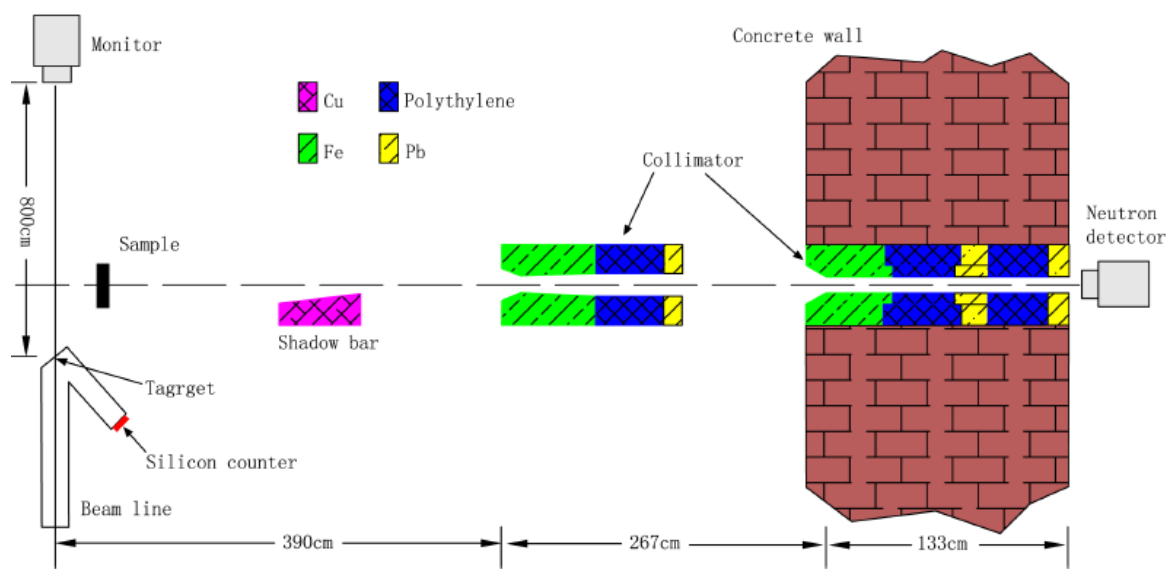

**Fig. 1**. (Color online) A schematic view of the experimental arrangement

A simplified block diagram of the electronic circuit for the present experiment is shown in Fig. 2. The dynode signal of neutron detector was amplified with ORTEC 590 amplifier and sent into analog-to-digital converter (ADC) for recording the pulse height. A pulse shape discrimination module of 2160A was used to obtain the pulse shape. The time of flight (TOF) was measured with a time-to-amplitude converter (TAC). Anode signal of neutron detector and delayed beam pick-up signal were used as start and stop signal of the TAC, respectively. The TAC calibration (~0.234 ns/channel) was made several times using a time calibrator module during the experiment. The TOF of the monitor and the pulse height of Silicon counter were also recorded. All experimental data were collected on an event-by-event basis using a CAMAC-based online data acquisition system.

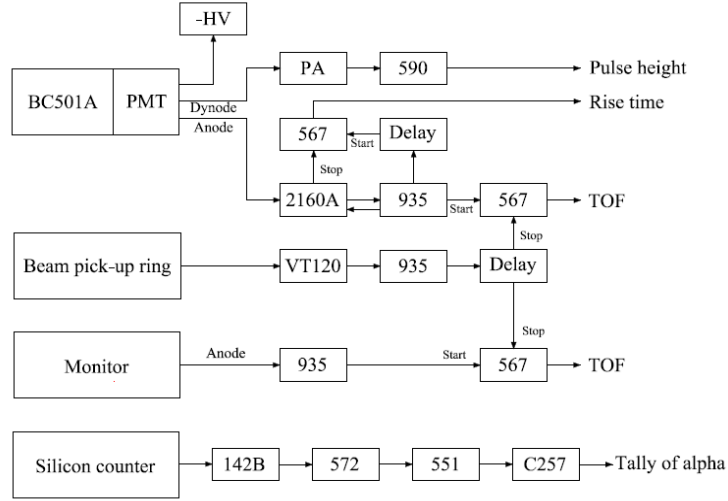

**HV**: High voltage, **PA**: Preamplifier, **590**: Amplifier, **567**: Time-to-amplitude converter, **2160A**: Pulse shape discrimination module, **935**: Constant fraction discriminator, **Delay**: Delay line, **VT120**: Fast timing preamplifier, **Monitor**: BC501A liquid scintillator, **142B**: Preamplifier, **572**: Amplifier, **551**: Timing single-channel analyzer, **C257**: Scaler.

**Fig. 2**. (Color online) Block diagram of the electronic circuit for the present experiment

## 2.2 Data analysis

The leakage neutron spectra were obtained from the results of the sample-in measurement by subtracting those of sample-out without gamma-ray events, and normalized by the number of source neutrons. The source neutron yield at the target position can be estimated by the following equation (1):

$$N_n = \frac{N_\alpha \times A_\alpha \times \sigma_{tot}}{\Delta\Omega \times \sigma(\theta)} \qquad (1)$$

where $N_n$ is the source neutron yield, $N_\alpha$ is the number of alpha particles, $\Delta\Omega$ is the solid angle of silicon counter, $A_\alpha$ is the anisotropic factor, which is relevant to average deuteron beam energy and emission angle of alpha particle, $\sigma_{tot}$ and $\sigma(\theta)$ are total and differential cross sections of the T (d, n) $^4$He reaction, respectively. Neutron events were separated from the gamma ones clearly by the pulse shape discrimination (PSD) method. The PSD was performed in the offline analysis using the two-dimensional distributions of the rise time versus pulse height of the neutron detector. Neutrons appear in higher rise time channels than gamma-rays for having pulses with a long fluorescence tail. The absolute flight time from the sample to the detector was calibrated using a peak of elastic scattering neutrons at the TOF spectrum. The energy spectrum of the neutrons is calculated from the TOF spectrum with the neutron detection efficiency. The light output function and the neutron detection efficiency were well calibrated using $^9$Be (d, n) $^{10}$B neutron source at HI-13 Tandem Accelerator at CIAE. With the known light output function, the detection efficiency was obtained from the calculated results of NEFF Monte Carlo code [21]. The uncertainties of the present experiment were due to statistical and systematic errors. The systematic errors were mainly caused by neutron detection efficiency (about 5%), and source neutron yield (about 3%).

Fig. 3 shows the measured leakage neutron TOF spectrum for 14.8 MeV neutrons on a polyethylene sample at 60°. The measured spectrum is compared with the MCNP calculated one with the ENDF/B-VII.0 evaluated nuclear data. The comparison shows that the experimental result is well reproduced by the calculated one in the whole neutron energy range. This result indicates that the experimental apparatus and the data analyzing procedures are in reasonable shape.

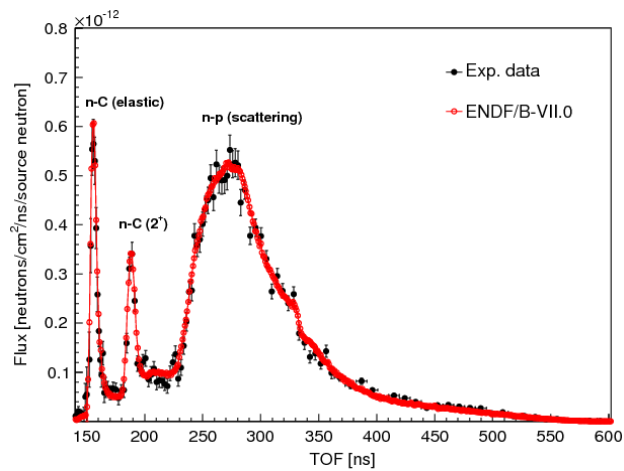

**Fig. 3.** (Color online) Leakage neutron spectrum for 14.8 MeV neutrons on a polyethylene sample at 60°

## 3. Monte Carlo Calculation

MCNP is a general-purpose Monte Carlo N-Particle code developed by Los Alamos National Laboratory for simulating the transport of neutron, photon and electron though matter. It is widely used in designing facilities of engineering applications related to neutrons, because of the outstanding features, including the capability to calculate $k_{eff}$ eigenvalues for critical systems.

In the present work, the leakage neutron spectra for a Tungsten slab sample with the size of 10 × 10 × 7 cm$^3$ were simulated by MCNP-4C code. A detailed model of the experimental target setup was described in the simulation and employed to define the value of source CEL variable where the particle started. The SDEF card (general source) was utilized for the source specification. Both of the angular distribution and angle dependent energy distribution of the source neutrons were calculated by TARGET code [22] with the experimental deuteron beam condition and the T-Ti target setup. The value of TME variable was defined by using experimentally measured time of flight spectra of source neutrons by the monitor. A point detector estimator was used to tally the leakage neutron time of flight spectra for comparing with the experimentally measured ones.

**4. Results and Discussion**

The measured leakage neutron spectra for the Tungsten sample at 60° and 120° are shown in Fig. 4 comparing with the calculated ones using the five evaluated nuclear data libraries. The calculation-to-experiment (C/E) values of the spectra integrated over five energy regions are given in Fig. 5 and Table 1. From these results, the following observations are made:

(1) In the 13.5-16 MeV neutron energy range, the major contribution in the neutron spectrum comes from the elastic scattering. The calculated elastic scattering peak with the evaluated data of the ENDF/B-VII.0 and JENDL-4.0 libraries are higher than the experimental one, while those of the ADS-2.0, ENDF/B-VII.1 and CENDL-3.1 libraries are underestimated. It turns out that these disagreements are angular dependent and became larger at 120°.

(2) In the 8.5-13.5 MeV neutron energy range, the calculated spectra with the evaluated data of ENDF/B-VII.1 give agreement with the experimental ones within 8% and 12% of discrepancies at 60° and 120°, respectively. The calculated spectra with ADS-2.0 show better agreements within 2% at 120° and 5% at 60°. While, the calculated spectra with ENDF/B-VII.0, JENDL-4.0 and CENDL-3.1 are largely underestimated, especially at 60°.

(3) In the 4-8.5 MeV neutron energy range, the results from ADS-2.0 and ENDF/B-VII.1 agree with the experimental ones in less than 14% of discrepancies. The calculated spectra with the ENDF/B-VII.0 library show a much better agreement with the experimental ones in less than 5% of discrepancies at 60°, but significantly overpredict by 29% at 120°. The predictions based on both of JENDL-4.0 and CENDL-3.1 are worse than those of the other three libraries in this energy region.

(4) Below 4 MeV, the calculated spectra with the ADS-2.0, ENDF/B-VII.0, ENDF/B-VII.1 and JENDL-4.0 libraries reproduce the experimental ones within 7% of discrepancies, while those with the CENDL-3.1 library overestimate by 16%.

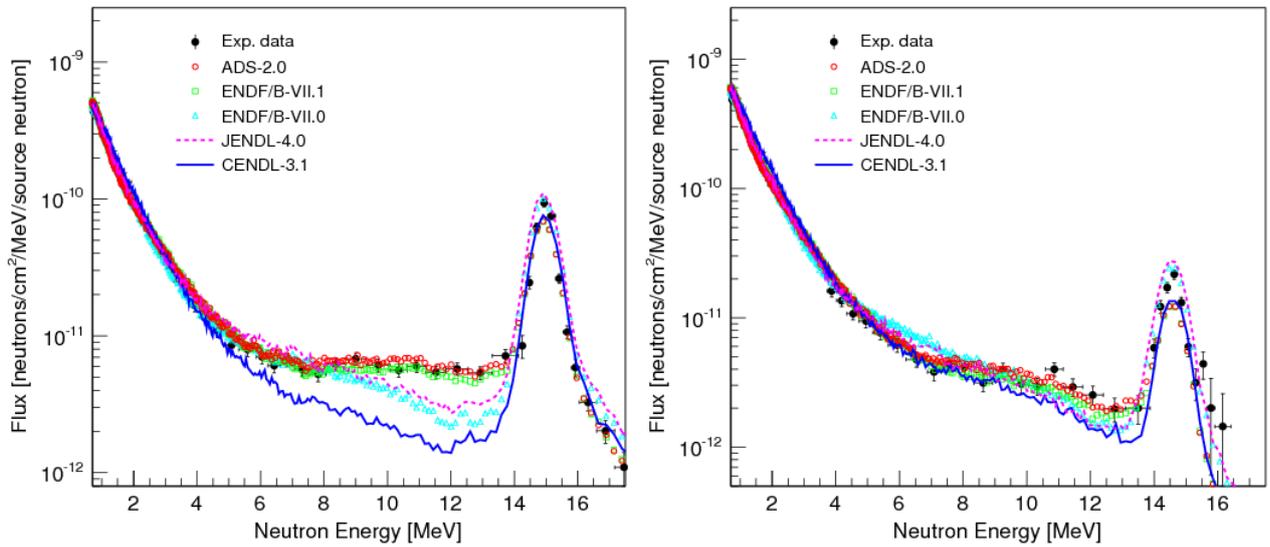

**Fig. 4.** (Color online) Comparison of experimental and calculated neutron spectra for thickness of 7 cm at 60° (left) and at 120° (right)

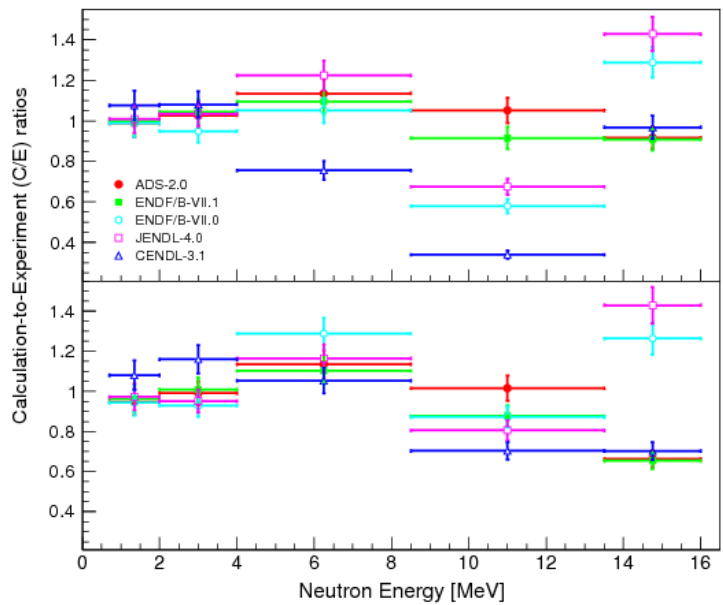

**Fig. 5.** (Color online) The C/E values integrated over the five energy regions for thickness of 7cm at 60° (top) and at 120° (bottom)

Table 1 The C/E values of the spectra integrated over 5 energy regions for
7 cm thickness at 60° and 120°

| | Energy (MeV) | C/E (ADS-2.0) | C/E (ENDF/B-VII.1) | C/E (ENDF/B-VII.0) | C/E (JENDL-4.0) | C/E (CENDL-3.1) |
|---|---|---|---|---|---|---|
| 60° | 0.7–2 | 0.987 ±0.068 | 1.000 ±0.068 | 0.987 ±0.068 | 1.009 ±0.069 | 1.076 ±0.074 |
| | 2–4 | 1.026 ±0.062 | 1.044 ±0.062 | 0.950 ±0.057 | 1.035 ±0.062 | 1.080 ±0.065 |
| | 4–8.5 | 1.135 ±0.068 | 1.095 ±0.066 | 1.051 ±0.063 | 1.224 ±0.073 | 0.755 ±0.045 |
| | 8.5–13.5 | 1.051 ±0.062 | 0.915 ±0.054 | 0.579 ±0.034 | 0.675 ±0.040 | 0.340 ±0.020 |
| | 13.5–16 | 0.916 ±0.054 | 0.908 ±0.054 | 1.288 ±0.076 | 1.428 ±0.085 | 0.968 ±0.058 |
| 120° | 0.7–2 | 0.947 ±0.065 | 0.963 ±0.066 | 0.944 ±0.064 | 0.972 ±0.066 | 1.080 ±0.074 |
| | 2–4 | 0.991 ±0.059 | 1.010 ±0.061 | 0.928 ±0.056 | 0.951 ±0.057 | 1.160 ±0.070 |
| | 4–8.5 | 1.135 ±0.068 | 1.103 ±0.066 | 1.288 ±0.078 | 1.164 ±0.070 | 1.053 ±0.064 |
| | 8.5–13.5 | 1.016 ±0.062 | 0.877 ±0.054 | 0.872 ±0.054 | 0.805 ±0.050 | 0.703 ±0.044 |
| | 13.5–16 | 0.663 ±0.042 | 0.654 ±0.042 | 1.263 ±0.080 | 1.429 ±0.091 | 0.702 ±0.045 |

In order to find out the reason for the discrepancies between the calculated spectra with the different evaluated data and the measured ones, the total and several partial cross sections in the evaluated nuclear data libraries are studied. The emission neutrons of the (n, el), (n, inl), (n, 2n), (n, 3n) and (n, np) reaction channels make major contributions to the leakage neutron spectra of the present work. The angular distributions of the neutron elastic scattering for Tungsten at the incident neutron energy of 15 MeV in the JENDL-4.0 and ENDF/B-VII.1 libraries are shown in Fig. 6. The JENDL-4.0 cross sections are higher than the ENDF/B-VII.1 ones at both 60° and 120°. This may explain the differences in the elastic scattering peak of the neutron spectrum. The contributions to the total energy spectra from the continuum inelastic and (n, 2n) for Tungsten at incident neutron energy of 15 MeV in CENDL-3.1 and ENDF/B-VII.0 libraries are shown in Fig. 7. The contribution of the (n, 2n) reaction in the CENDL-3.1 is considerably higher than ENDF/B-VII.0 below 4 MeV and lower between 4 to 8 MeV, where the calculations with CENDL-3.1 give significant disagreement with the experimental ones. The contribution from the continuum inelastic scattering in CENDL-3.1 is significantly lower than that in ENDF/B-VII.0 between 8 to 13.5 MeV, which may cause the discrepancies between the measured spectra and the calculated ones at this energy range.

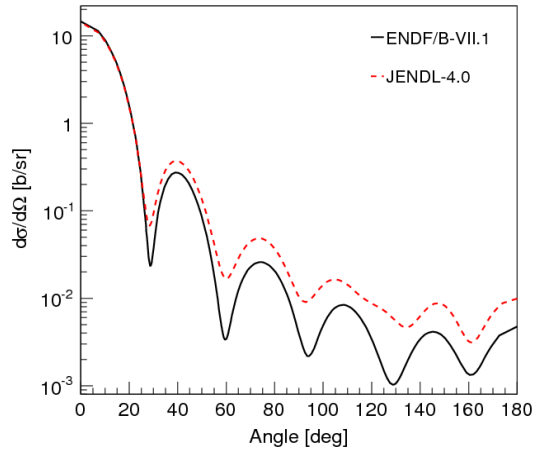

**Fig. 6.** (Color online) The angular distributions of the neutron elastic scattering for Tungsten at the incident neutron energy of 15 MeV in JENDL-4.0 and ENDF/B-VII.1

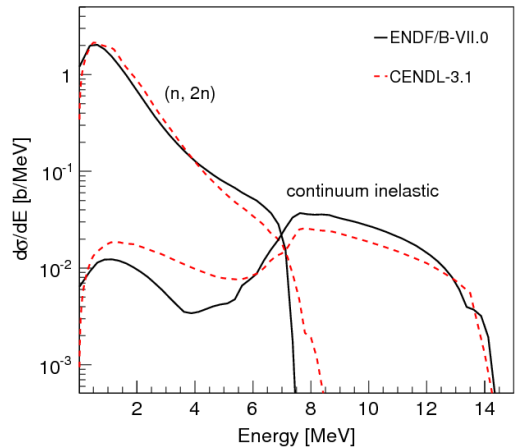

**Fig. 7.** (Color online) The contributions to the total energy spectra from the continuum inelastic and (n, 2n) for Tungsten at incident neutron energy of 15 MeV in CENDL-3.1 and ENDF/B-VII.0

## 5. Conclusion

The validations of the evaluated nuclear data for Tungsten are performed, using the leakage neutron spectra experimentally measured by a time-of-flight method. The theoretical calculations were carried out by MCNP-4C Monte Carlo code using the evaluated nuclear data of the ADS-2.0, ENDF/B-VII.0, ENDF/B-VII.1, JENDL-4.0 and CENDL-3.1 libraries. From the comparisons, it is found that the calculations with the evaluated data of the ADS-2.0 and ENDF/B-VII.1 libraries fairly well reproduce the experimental results within 14% and 10% of discrepancies in the whole energy range at 60°. However, a significant disagreement is observed at 120° in the elastic scattering peak. The calculated results with the ENDF/B-VII.0, JENDL-4.0 and CENDL-3.1 libraries give agreements with the experimental ones in less than 16% below 4 MeV, while strong discrepancies are observed in the 8.5-13.5 MeV and in the elastic scattering peak. These discrepancies may be due to the improper evaluation of the angular distribution and secondary neutron energy distribution of the elastic scattering, inelastic scattering and (n, 2n) reaction channels in evaluated nuclear data libraries. In overall, the calculations with the ADS-2.0 and ENDF/B-VII.1 libraries give satisfactory reproductions of the experimentally measured neutron leakage spectra in this study, although a slight adjustment is needed for the angular distribution of the elastic scattering at $120°$. Those with the ENDF/B-VII.0, JENDL-4.0, and CENDL-3.1 result in larger discrepancies, especially in the inelastic contributions at the 8.5-13.5 MeV neutron energy range.

The present work also shows that the experimental apparatus and the data analysis procedures work well and provide valuable data for benchmarking the evaluated nuclear data for Tungsten. The further integral benchmark experiments and differential cross section experiments will be investigated for materials related to design of the CIADS in the future in the same frame of this work.